\documentclass[prd,showpacs,superscriptaddress,preprintnumbers,
               nofootinbib,amsmath,amssymb]{revtex4}
\usepackage[dvips]{graphicx}

\begin{document}
\title{Chiral effective model with the Polyakov loop}
\author{Kenji Fukushima}
\email{kenji@lns.mit.edu}
\affiliation{Center for Theoretical Physics,
             Massachusetts Institute of Technology,
             Cambridge, Massachusetts 02139}
\affiliation{Department of Physics, University of Tokyo,
             7-3-1 Hongo, Bunkyo-ku, Tokyo 113-0033, Japan}

\begin{abstract}
We discuss how the simultaneous crossovers of deconfinement and chiral
restoration can be realized. We propose a dynamical mechanism assuming
that the effective potential gives a finite value of the chiral
condensate if the Polyakov loop vanishes. Using a simple model, we
demonstrate that our idea works well for small quark mass, though
there should be further constraints to reach the perfect locking of
two phenomena.
\end{abstract}
\pacs{11.10.Wx, 11.30.Rd, 12.38.Aw, 25.75.Nq}
\preprint{MIT-CTP 3424}
\maketitle


\paragraph*{\bf Introduction}
Chiral symmetry plays an important role in effective model approaches
to Quantum Chromodynamics (QCD). The hadronic properties at low energy
have been successfully described by chiral effective models such as
the linear sigma model \cite{lee72}, the Nambu--Jona-Lasinio (NJL)
model \cite{nam61,kle92}, the chiral random matrix model \cite{shu93},
chiral perturbation theory \cite{wei79} and so on. The nature of the
chiral phase transition at finite temperature can be classified by
chiral symmetry according to the universality argument \cite{pis84}
and investigated in these effective models \cite{mey96}. In
particular, in the massless two-flavor case, we can expect a chiral
phase transition of second-order that belongs to the same universality
class as the 3d O(4) spin model. Then, we can anticipate what could
occur in the real world with finite but sufficiently small (up and
down) quark masses.

The deconfinement phase transition is rather obscure and veiled
because it is only well-defined in the heavy quark limit, which is too
far from the real world. In the heavy quark limit, i.e., in the
absence of dynamical quarks, the Polyakov loop serves as an order
parameter for deconfinement and the phase transition is characterized
by the spontaneous breaking of center symmetry \cite{pol78}. In the
presence of dynamical quarks the center symmetry is explicitly broken
and no order parameter or criterion has been established for the
deconfinement transition \cite{fuk03}. As mentioned above, on the
other hand, the chiral phase transition at high temperature or baryon
density has been well-understood by means of effective models. Those
model studies based on chiral symmetry, however, lack any dynamics
coming from the Polyakov loop, except for some efforts to clarify the
interplay between chiral dynamics and the Polyakov loop
\cite{goc85,ilg85,dig01,fuk03_2,san03}.

It is often argued that the \textit{mixing} between the Polyakov loop,
$L$, and the chiral order parameter (chiral condensate), $\chi$, can
account for the lattice QCD observation that deconfinement and chiral
restoration occur at the same pseudo-critical temperature
\cite{kar94}. The mixing argument is, however, not sufficient to give
a satisfactory explanation on the lattice QCD data. In the ($T,\,m_q$)
plane ($m_q$ being the current quark mass), as discussed in
\cite{hat03}, there appear two terminal points of the first-order
phase boundary, namely, the critical end-points (CEPs). One is the
chiral CEP denoted by ($T_{\textsf{C}},\,m_q^{\textsf{C}}$) and the
other is the deconfinement CEP denoted by
($T_{\textsf{D}},\,m_q^{\textsf{D}}$). For example, in the two-flavor
three-color case, it is known that
($T_{\textsf{C}}\sim170\;\text{MeV},\,m_q^{\textsf{C}}=0$) and
($T_{\textsf{D}}\sim270\;\text{MeV},\,m_q^{\textsf{D}}\sim1\;\text{GeV}$)
\cite{has83,fuk03_2}. It should
be noted that the CEP is a true second-order critical point with a
\textit{divergent susceptibility}. Mixing means that $L$ and $\chi$
should share the \textit{same} singularity in their susceptibilities
near the CEP; for $m_q=m_q^{\textsf{C}}$ (and $m_q^{\textsf{D}}$), the
susceptibilities of $L$ and $\chi$ both diverge at $T=T_{\textsf{C}}$
(and $T_{\textsf{D}}$ respectively). The susceptibility peak is
smeared as $m_q$ leaves from the CEP. The important point is that,
with $m_q$ fixed at a certain value near the CEP, \textit{a smeared
bump originating from the other CEP may be observed separately as well
as a sharp peak from the closer CEP}. Thus the mixing argument cannot
exclude a double-peak structure. For $m_q\simeq m_q^{\textsf{C}}$ for
example, the Polyakov loop susceptibility can have a sharp peak around
$T_{\textsf{C}}$ (coming from the mixing with the diverging chiral
susceptibility) as well as a broad bump around $T_{\textsf{D}}$
(coming from a remnant of deconfinement). We should be cautious about
the mixing argument to understand the lattice QCD data in which no
double-peak structure has been seen for any $m_q$ \cite{kar94,hat03}.

In fact, the \textit{locking} between the two crossover phenomena
depends on the detailed properties of the interaction. The purpose of
this letter is to propose a simple mechanism to exclude the
undesirable double-peak structure and to complement the shortcomings
of the mixing argument.
\vspace{1mm}


\paragraph*{\bf Idea}
To make our idea clear in general setting, let us suppose that we have
a \textit{full} effective potential, i.e.,
$V_{\text{eff}}[L,\chi;m_q]$. In principle, the behavior of $L$ and
$\chi$ should be completely determined by
$V_{\text{eff}}[L,\chi;m_q]$. Thus the question arises; what property
of  $V_{\text{eff}}[L,\chi;m_q]$ can give rise to the simultaneous
crossovers? Our idea is as follows.

First of all, an important property follows from the theoretical
arguments given by Casher \cite{cas79} and 't~Hooft \cite{tho80}.
According to their arguments, the confined phase must have a
non-vanishing chiral condensate, which suggests that the chiral phase
transition should occur at higher temperature than deconfinement. This
means that $V_{\text{eff}}[L=0,\chi;m_q=0]$ leads to $\chi\neq0$ at
any temperature if $L=0$ is imposed by hand (or approximately chosen
as a minimum of the effective potential). This property has not been
proven in QCD (see also \cite{ito83}) but is realized in the strong
coupling analysis \cite{fuk03_2} and assumed here.

Next, because $L$ has turned out to behave approximately as an order
parameter in lattice simulations \cite{kar94}, we can expect that $L$
is almost zero below the deconfinement crossover temperature,
$T_{\text{d}}$, regardless of dynamical quarks. Then, together with
the above property, $\chi$ must have a non-vanishing value below
$T_{\text{d}}$ even for $m_q\sim0$ (i.e., spontaneous chiral symmetry
breaking). Thus the chiral restoration temperature, $T_\chi$, is
greater than or equal to $T_{\text{d}}$.

In contrast, the critical temperatures at the CEPs are
$T_{\textsf{D}}\simeq270\;\text{MeV (for $m_q=\infty$)}>
 T_{\textsf{C}}\simeq170\;\text{MeV (for $m_q=0$)}$, which implies
$T_{\text{d}}\ge T_\chi$ for an intermediate value of $m_q$, unless
chiral or center symmetry is overwhelmingly broken.

Our idea is that $T_{\text{d}}=T_\chi$ is likely to be realized by
$T_\chi\ge T_{\text{d}}$ from the properties of the effective
potential and $T_{\text{d}}\ge T_\chi$ from, so to speak, the boundary
condition. Chiral symmetry is broken by $m_q\neq0$, while the center
symmetry breaking is suppressed by the constituent quark mass even for
small $m_q$. Hence, our idea is expected to work especially for
$m_q\sim0$. This mechanism can complement the mixing argument and lead
to a robust single-peak structure in the susceptibilities.
\vspace{1mm}


\paragraph*{\bf Effective Model}
For the purpose of demonstrating our idea, we propose a simple chiral
effective model with Polyakov loop dynamics. If the Polyakov gauge
($A_4$ is static and diagonal) is employed, the Polyakov loop (or
strictly speaking its phase)  appears in the quark action as an
imaginary quark chemical potential \cite{wei81,fuk00}. Thus we can
uniquely determine the coupling between the Polyakov loop and quark
excitations. We shall take this advantage by adopting the NJL model
which is given in terms of quark degrees of freedom.

The conventional Lagrangian density of the NJL model is
\begin{equation}
 \mathcal{L}_{\text{NJL}}=
  \bar{q}\bigl(\mathrm{i}\gamma^\mu \partial_\mu-m_q\bigr)q
  +\frac{G}{2}\bigl\{(\bar{q}q)^2+(\bar{q}\mathrm{i}\gamma_5
  \vec{\tau}q)^2\bigr\},
\end{equation}
where $m_q=5.5\,\text{MeV}$ and $G=2\times 5.496\,\text{GeV}^{-2}$.
The momentum integration is regulated by the cut-off
$\Lambda=631\,\text{MeV}$. These model parameters are chosen as to
reproduce the pion mass and decay constant at zero temperature
\cite{kle92}. In the mean field approximation the thermodynamic
potential is given by
\begin{equation}
 \Omega_{\text{NJL}}/V=\frac{1}{2G}(M-m_q)^2-2N_{\text{c}}
  N_{\text{f}}\int\frac{\mathrm{d}^3 p}{(2\pi)^3}\Bigl\{E_p
  +T\ln\bigl[1+\mathrm{e}^{-(E_p-\mu)/T}\bigr]
  +T\ln\bigl[1+\mathrm{e}^{-(E_p+\mu)/T}\bigr]\Bigr\}.
\end{equation}
$V$ is the spatial volume, $N_{\text{f}}$ is the flavor number fixed
as $N_{\text{f}}=2$ throughout this letter, and $\mu$ is the quark
chemical potential. We neglect any $\mu$ dependence in $G$ as usual
\cite{kle92}. The energy of quasi-quarks is given by
$E_p=\sqrt{p^2+M^2}$ with the constituent quark mass
$M=m_q-G\langle\bar{q}q\rangle$. The cut-off is imposed only on the
first term in the curly brackets (zero-point energy) in the present
analysis. The finite temperature contribution has a natural cut-off in
itself specified by the temperature. Identifying the imaginary quark
chemical potential with the Polyakov  loop, we can define our model by
the following thermodynamic potential,
\begin{align}
 \Omega/V &=V_{\text{glue}}[L]+\frac{1}{2G}(M-m_q)^2 \notag\\
  &\quad -2N_{\text{c}}N_{\text{f}}\int
  \frac{\mathrm{d}^3 p}{(2\pi)^3} \Bigl\{E_p+T\frac{1}{N_{\text{c}}}
  \mathrm{Tr_c}\ln\Bigl[1+L\mathrm{e}^{-(E_p-\mu)/T}\Bigr]+T
  \frac{1}{N_{\text{c}}}\mathrm{Tr_c}\ln
  \Bigr[1+L^\dagger\mathrm{e}^{-(E_p+\mu)/T}\Bigr]\Bigr\},
\label{eq:potential}
\end{align}
where the Polyakov loop is an SU($N_{\text{c}}$) matrix in color space
explicitly given by
\begin{equation}
 L(\vec{x})=\mathcal{T}\exp\biggl[-\mathrm{i}\int_0^\beta
  \mathrm{d}x_4\, A_4(x_4,\vec{x})\biggr].
\end{equation}
This coupling between the Polyakov loop and the chiral condensate can
be derived also in the strong coupling approach \cite{ilg85,fuk03_2}.
The generalization to aQCD (QCD with dynamical quarks in the adjoint
representation in color space) is readily available if one replaces
the Polyakov loop by the Polyakov loop in the adjoint representation
(c.f.\ \cite{fuk03_2}). In this letter we will focus only on the case
in the fundamental representation.

Since the four-quark coupling constant, $G$, contains the information
on gluons, $G$ should depend on $L$. We simply neglect this possible
$L$ dependence. Nevertheless, this $L$ dependence makes no qualitative
difference because $G$ incorporates all gluons and would not be much
affected by $L$ only, where $L$ is essentially the temporal component
of gluons. This approximation is acceptable in the same level as
neglecting the possible $\mu$ dependence in $G$.

It is worth noting that the NJL model with both quarks and the
Polyakov loop is not incompatible with confinement at low temperature.
To make this clear, let us assume that confinement corresponds to the
condition, $L=0$, though this is not precise due to the center
symmetry breaking. The Taylor expansion of the logarithmic terms
generates a series in powers of $L$ whose power corresponds to the
quark excitation number. $L, L^2, L^4,\dots$ terms have non-trivial
triality and break the center symmetry, while $L^3, L^\dagger
L(=1),\dots$ terms have zero triality (color-singlet part)
corresponding to baryons, mesons, etc. In a more elaborated
approximation scheme used in \cite{fuk03,fuk03_2}, the group
integration of $L$ with a vanishing mean-field, $L=0$, singles out
only the terms with zero triality. [Our notation is sloppy as long as
no confusion arises. $L$ is used for both the Polyakov loop matrix and
its traced expectation value.] Roughly speaking, the limit of $L=0$
allows only excitations with zero triality such as baryons and mesons.
In the present treatment the same argument holds in part (see
(\ref{eq:ansatz})). Although our model (\ref{eq:potential}) is
described in terms of colored quarks, the thermal excitation must
consist of colorless composites as long as $L$ is small. Once $L$ gets
larger at higher temperature, our model would reduce into the standard
NJL model allowing thermal quark excitations.

The remaining part,  $V_{\text{glue}}[L]$, is the effective potential
only in terms of the Polyakov loop. Here we shall adopt a simple
choice for $V_{\text{glue}}[L]$ at the sacrifice of quantitative
accuracy; we employ the leading order result of the strong coupling
expansion that is simple and yet reasonable qualitatively as compared
with the lattice results \cite{pol82}. This potential has only one
parameter $a$ (the lattice spacing). For $N_{\text{c}}=3$ we can write
it as
\begin{equation}
 V_{\text{glue}}[L]\cdot a^3/T=-2(d-1)\,\mathrm{e}^{-\sigma a/T}
  \bigl|\mathrm{Tr_c}L\bigr|^2
 -\ln\Bigl[-\bigl|\mathrm{Tr_c}L\bigr|^4+8\text{Re}\bigl(\mathrm{Tr_c}
  L\bigr)^3-18\bigl|\mathrm{Tr_c}L\bigr|^2+27\Bigr]
\end{equation}
with the string tension $\sigma=(425\,\text{MeV})^2$. The first term
comes from the kinetic part and the second term is just the logarithm
of the Haar measure associated with the SU(3) group integration.
$V_{\text{glue}}[L]$ leads to a first order phase transition with the
critical coupling
$2(d-1)\mathrm{e}^{-\sigma a/T_{\text{d}}}=0.5153$. We can fix the
deconfinement transition temperature as the empirical value
$T_{\text{d}}=270\,\text{MeV}$ by choosing $a^{-1}=272\,\text{MeV}$.

Apparently, the present model has two cut-offs, i.e., $\Lambda$ and
$a^{-1}$. This means that the model has two independent scales for
chiral symmetry breaking and confinement, or for mesons and the
Polyakov loop. Since QCD has only one scale, $\Lambda$ and $a^{-1}$
should be related to each other in principle. Here we simply fix them
as model parameters since we are working in the effective model to
abstract the essence for each dynamics \cite{shu81}.
\vspace{2mm}


\begin{figure}
\includegraphics[width=6cm]{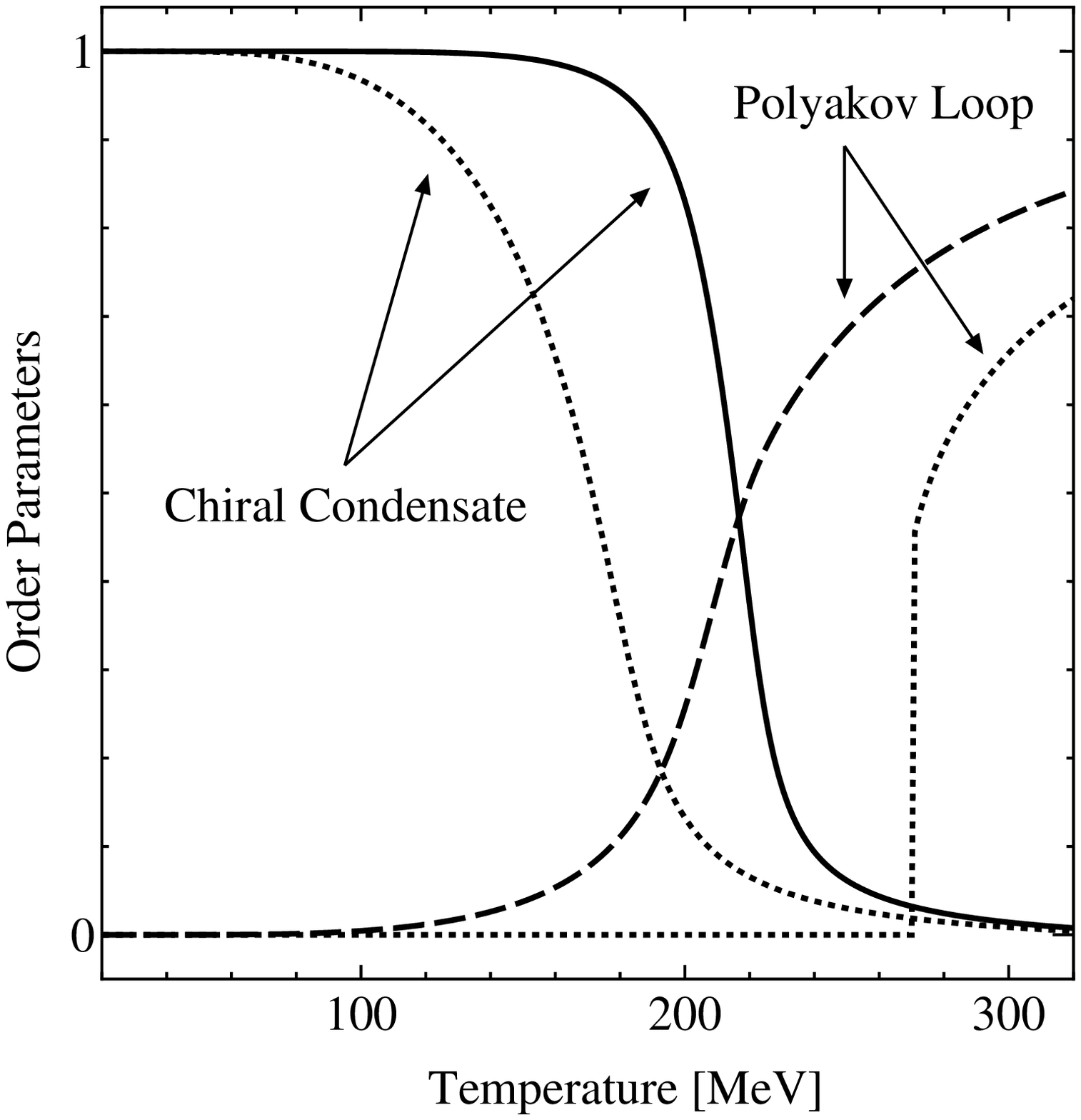} \hspace{9mm}
\includegraphics[width=6.8cm]{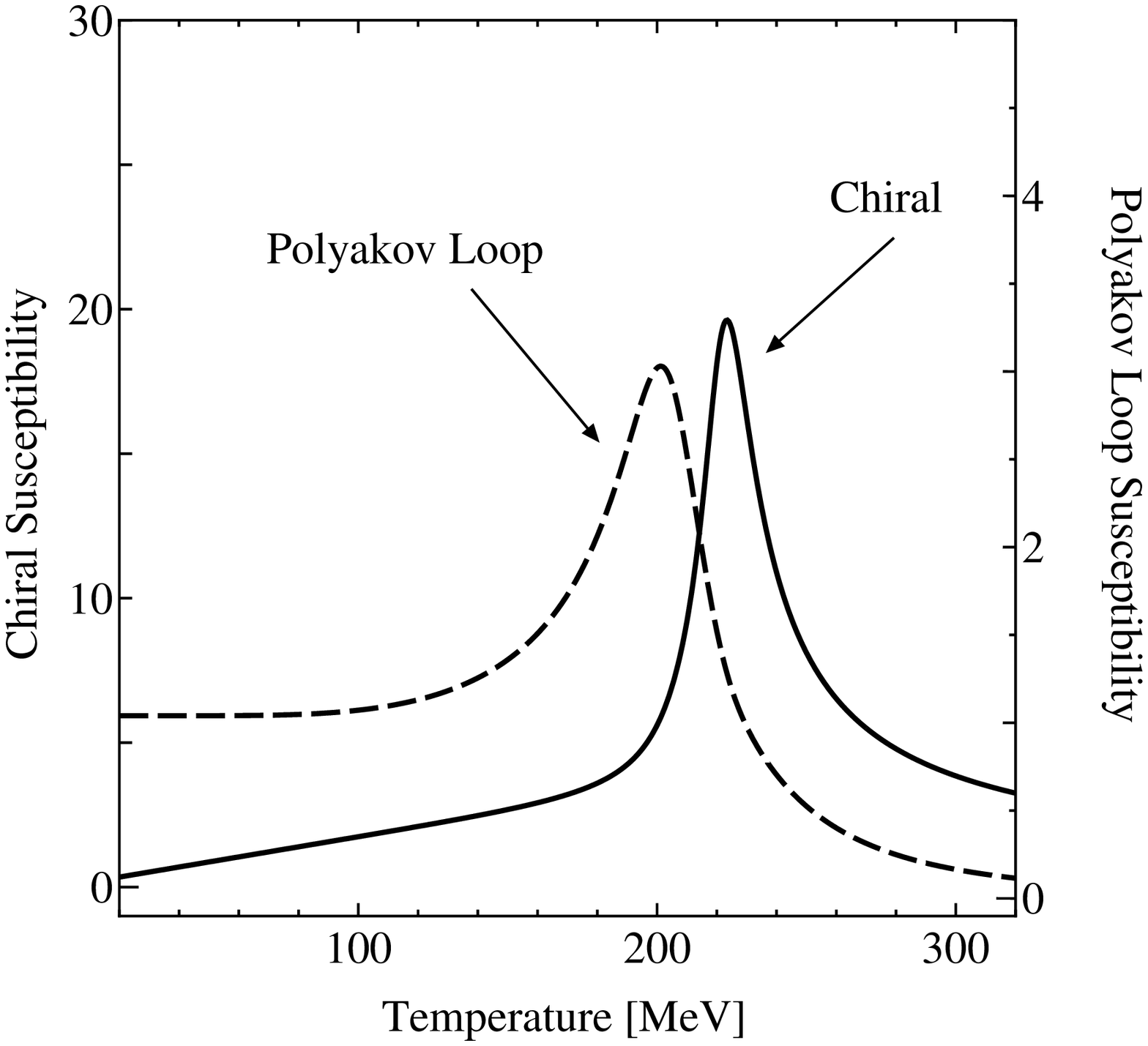}
\caption{The left figure is the behavior of the traced Polyakov loop
 $l$ and  the chiral condensate $\chi/\chi_0$ (normalized by the value
 at $T=0$) at $\mu=0$. The dotted curves represent $\chi/\chi_0$
 calculated from $\Omega_{\text{NJL}}$ and $l$ from $V_{\text{glue}}$.
 The right figure is the susceptibility.}
\label{fig:order}
\end{figure}

\paragraph*{\bf Numerical Results}
Before we refer numerical calculations in the $N_{\text{c}}=3$ case,
let us introduce an ansatz to simplify the analysis. In the Polyakov
gauge we can parametrize the SU(3) Polyakov loop matrix as
$L=\text{diag}(\mathrm{e}^{\mathrm{i}\phi},\mathrm{e}^{\mathrm{i}\phi'},
\mathrm{e}^{-\mathrm{i}(\phi+\phi')})$. The perturbative vacuum has
$\phi=\phi'=0$ and we can choose the confining vacuum at
$\phi=2\pi/3,\,\phi'=0$ \cite{wei81}. Thus we shall fix $\phi'=0$ from
the beginning for simplicity. Once this ansatz is accepted, we can
rewrite the potential (\ref{eq:potential}) only in terms of the traced
Polyakov loop, i.e.\ $l=(\mathrm{Tr_c}L)/N_{\text{c}}=(1+2\cos\phi)/3$.
After straightforward calculations we can reach the expression;
\begin{align}
 & \mathrm{Tr_c}\ln\Bigl[1+L\mathrm{e}^{-(E_p-\mu)/T}\Bigr]
  +\mathrm{Tr_c}\ln\Bigl[1+L^\dagger\mathrm{e}^{-(E_p+\mu)/T}\Bigr]
  \notag\\
 = & \ln\Bigl[1+(3\,l-1)\mathrm{e}^{-(E_p-\mu)/T}
  +\mathrm{e}^{-2(E_p-\mu)/T}\Bigr] + \ln\Bigl[1+(3\,l-1)
  \mathrm{e}^{-(E_p+\mu)/T}+\mathrm{e}^{-2(E_p+\mu)/T}\Bigr] \notag\\
 & \quad +\ln\Bigl[1+\mathrm{e}^{-(E_p-\mu)/T}\Bigr]
  \ln\Bigl[1+\mathrm{e}^{-(E_p+\mu)/T}\Bigr].
\label{eq:ansatz}
\end{align}
When $l=1$, the model is reduced into the standard two-flavor NJL
model having the chiral phase transition at $T_\chi=175\,\text{MeV}$
in the chiral limit. If $l$ is forced to be zero by hand, the
temperature effect is so suppressed that spontaneous chiral symmetry
breaking can sustain until $T_\chi\simeq 520\,\text{MeV}$, that is
much higher than $T_{\textsf{D}}$. Therefore the chiral phase
transition cannot occur until the Polyakov loop jumps from nearly zero
to a certain finite value. Our model (\ref{eq:potential}) satisfies
the essential assumption in our idea, $T_\chi \ge T_{\text{d}}$, for
$m_q\sim0$.

Figure \ref{fig:order} (left) shows the resulting behavior of order
parameters as functions of temperature at $\mu=0$. To see the
simultaneous crossovers clearly the chiral condensate
$\chi=\langle\bar{q}q\rangle$ is normalized by the value at $T=0$
(denoted by $\chi_0$). The results from $\Omega$ are represented by
the solid (for $\chi/\chi_0$) and dashed (for $l$) curves. The dotted
curves are the results from $\Omega_{\text{NJL}}$ and
$V_{\text{glue}}$ without any interaction between $\chi$ and $l$ for
reference. It should be noted that the mixing interaction between
$\chi$ and $l$ vanishes at zero temperature in this model.
Consequently the normalization $\chi_0$ is identical for both the
chiral condensates from $\Omega$ and $\Omega_{\text{NJL}}$.

In the presence of dynamical quarks, as seen from the figure, the
Polyakov loop shows a crossover around the pseudo-critical temperature
$T_{\text{c}}\simeq200\,\text{MeV}$. At the same time the chiral
condensate is affected by the Polyakov loop such that it tends to be
almost constant as long as $T<T_{\text{c}}$. The pseudo-critical
temperature can be read from the peak position of each susceptibility.
Here we shall define the dimensionless susceptibility of the chiral
order parameter and the Polyakov loop by using the curvature inferred
from the potential (\ref{eq:potential}). First, we define the
dimensionless curvature matrix $C$ by
\begin{equation}
 C_{qq} = \frac{\Lambda^2\partial^2}{\partial M^2}
  \frac{\beta\Omega}{\Lambda^3 V},\quad
 C_{ll} = \frac{\partial^2}{\partial l^2}
  \frac{\beta\Omega}{\Lambda^3 V}, \quad
 C_{ql} = C_{lq} = \frac{\Lambda\partial^2}{\partial M
  \partial l}\frac{\beta\Omega}{\Lambda^3 V}.
\end{equation}
Roughly speaking, $C^{-1}_{qq}$ ($C^{-1}_{ll}$) corresponds to the
chiral (Polyakov loop) fluctuation and $C_{ql}$ is the interaction
vertex of a quark and the Polyakov loop. Then, the susceptibility is
given by the inverse of $C$;
\begin{equation}
 \chi_q = \bigl(C^{-1}\bigr)_{qq}
  = \frac{C_{ll}}{C_{qq}C_{ll}-C_{ql}^2},\quad
 \chi_l = \bigl(C^{-1}\bigr)_{ll}
  = \frac{C_{qq}}{C_{qq}C_{ll}-C_{ql}^2}.
\end{equation}
The physical meaning of the above equations is transparent if we
notice that the fraction can be expanded as
$\chi_q=C^{-1}_{qq}+C^{-1}_{qq}C_{ql}C^{-1}_{ll}C_{lq}C^{-1}_{qq}+\cdots$,
that is the sum over mixing contributions. The mixing pattern is
similar to the relation between the chiral susceptibility and the
baryon number susceptibility \cite{kun00,fuj03}. As shown in
Fig.~\ref{fig:order} (right), our idea works pretty well so that the
peak of the Polyakov loop susceptibility can be found just near the
peak of the chiral susceptibility, though perfect coincidence is not
reached.

\begin{figure}
\includegraphics[width=6cm]{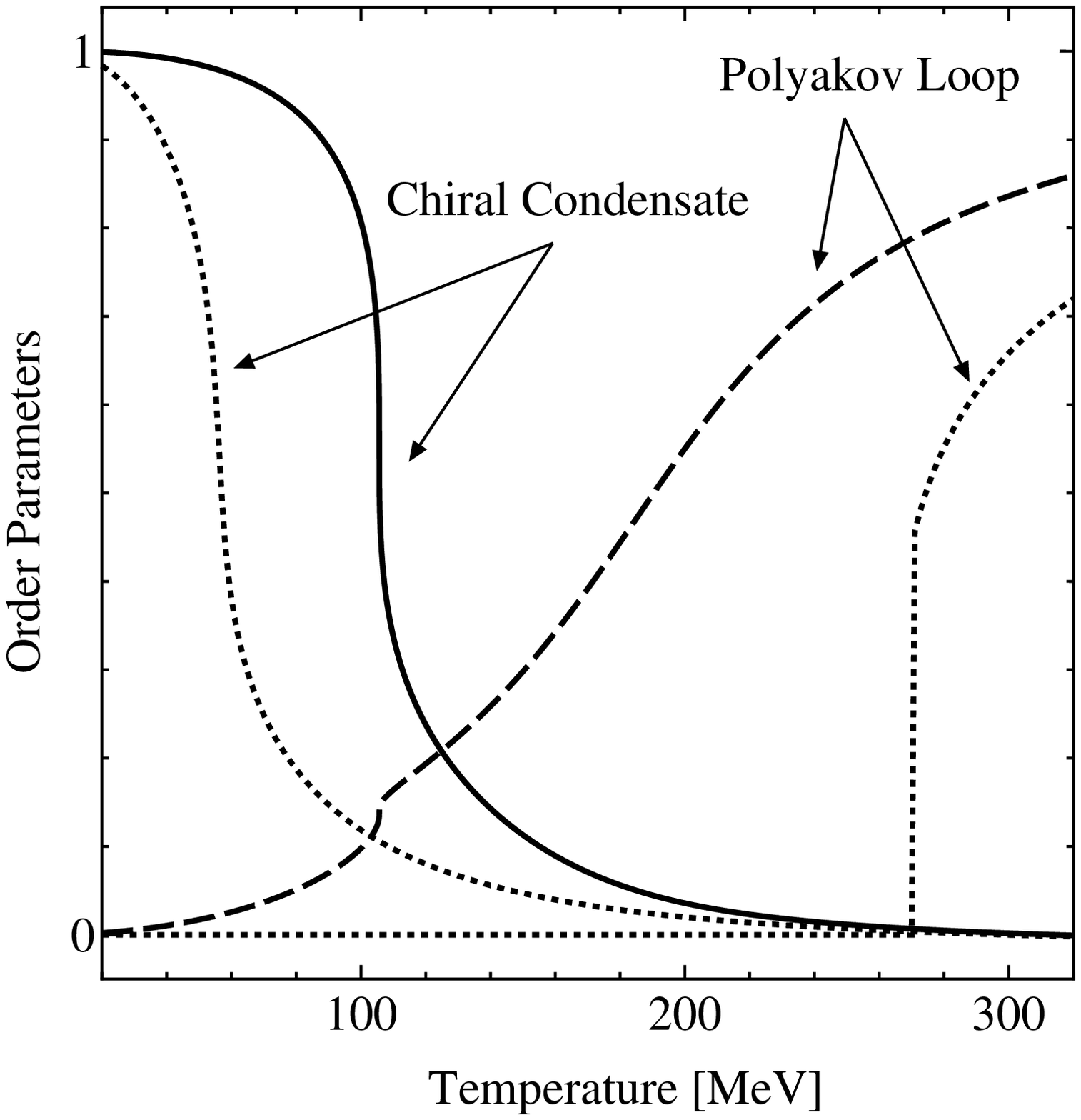} \hspace{9mm}
\includegraphics[width=6.8cm]{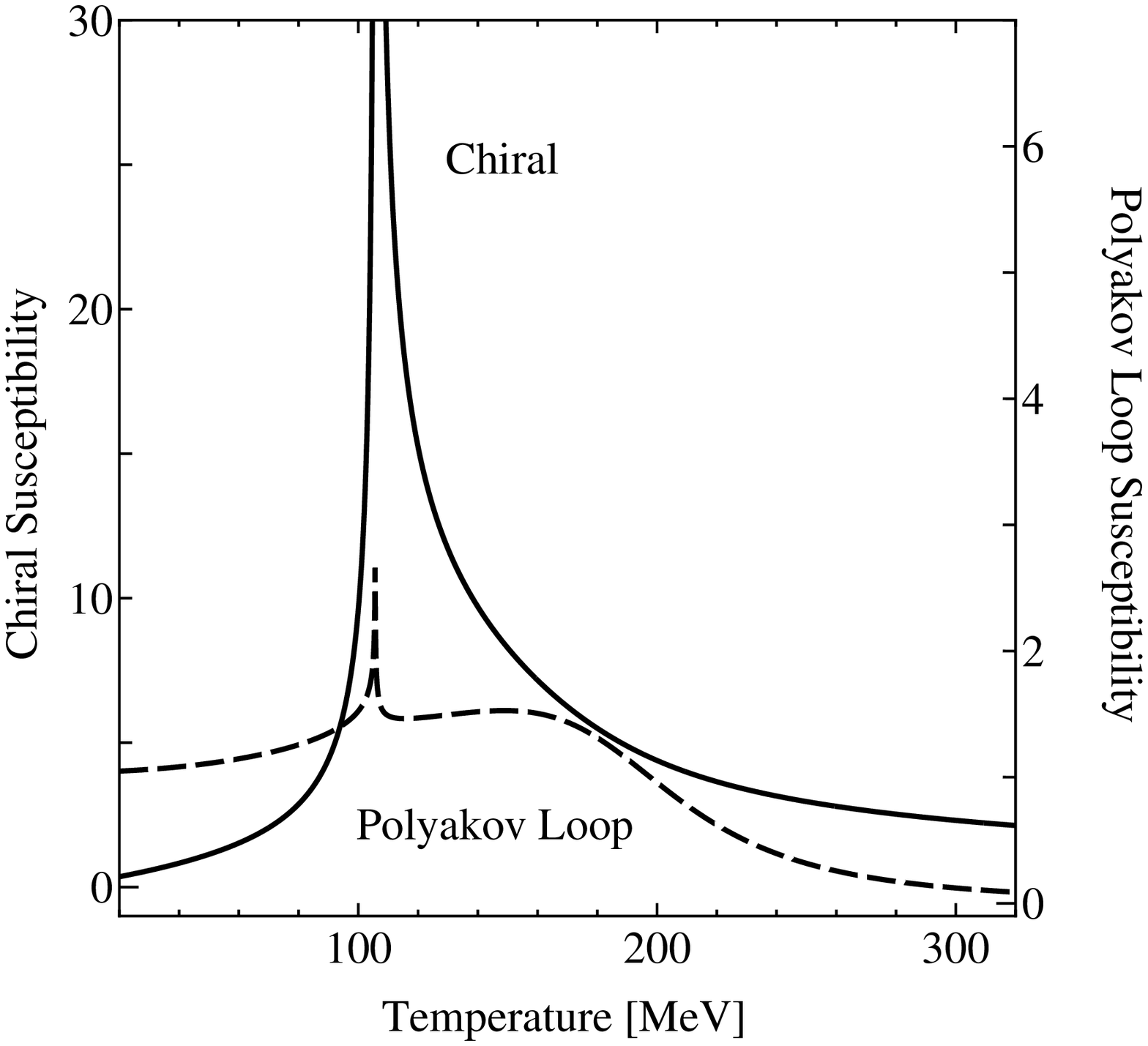}
\caption{The order parameter and susceptibility around the chiral
critical end-point with $\mu_{\textsf{E}}=321\,\text{MeV}$.}
\label{fig:order_ccep}
\end{figure}

When the quark chemical potential, $\mu$, becomes larger, the Polyakov
loop shows a crossover with smoother slope because the density effect
itself breaks the center symmetry explicitly. It is widely accepted
that the chiral phase transition becomes of first-order at large
$\mu$. In the ($\mu,\,T$) plane, therefore, we can expect another CEP
\cite{asa89,ste98}. Actually we found the CEP at
$(\mu_{\textsf{E}}=321\,\text{MeV},\,T_{\textsf{E}}=106\,\text{MeV})$,
which is close to the value originally obtained in \cite{asa89}. The
order parameter and the susceptibility around this CEP are shown in
Fig.~\ref{fig:order_ccep}.

\begin{figure}
\includegraphics[width=6cm]{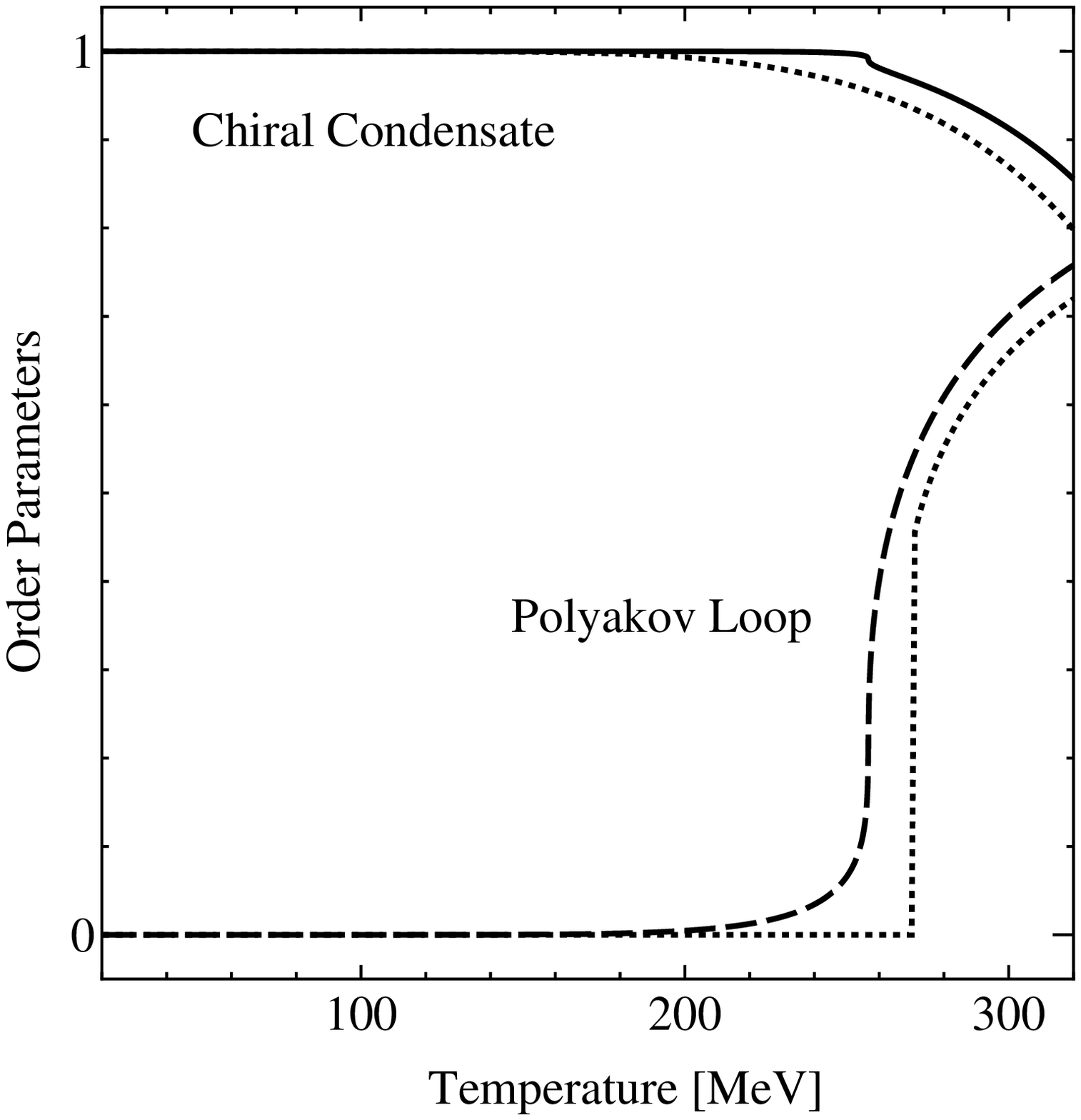} \hspace{9mm}
\includegraphics[width=6.8cm]{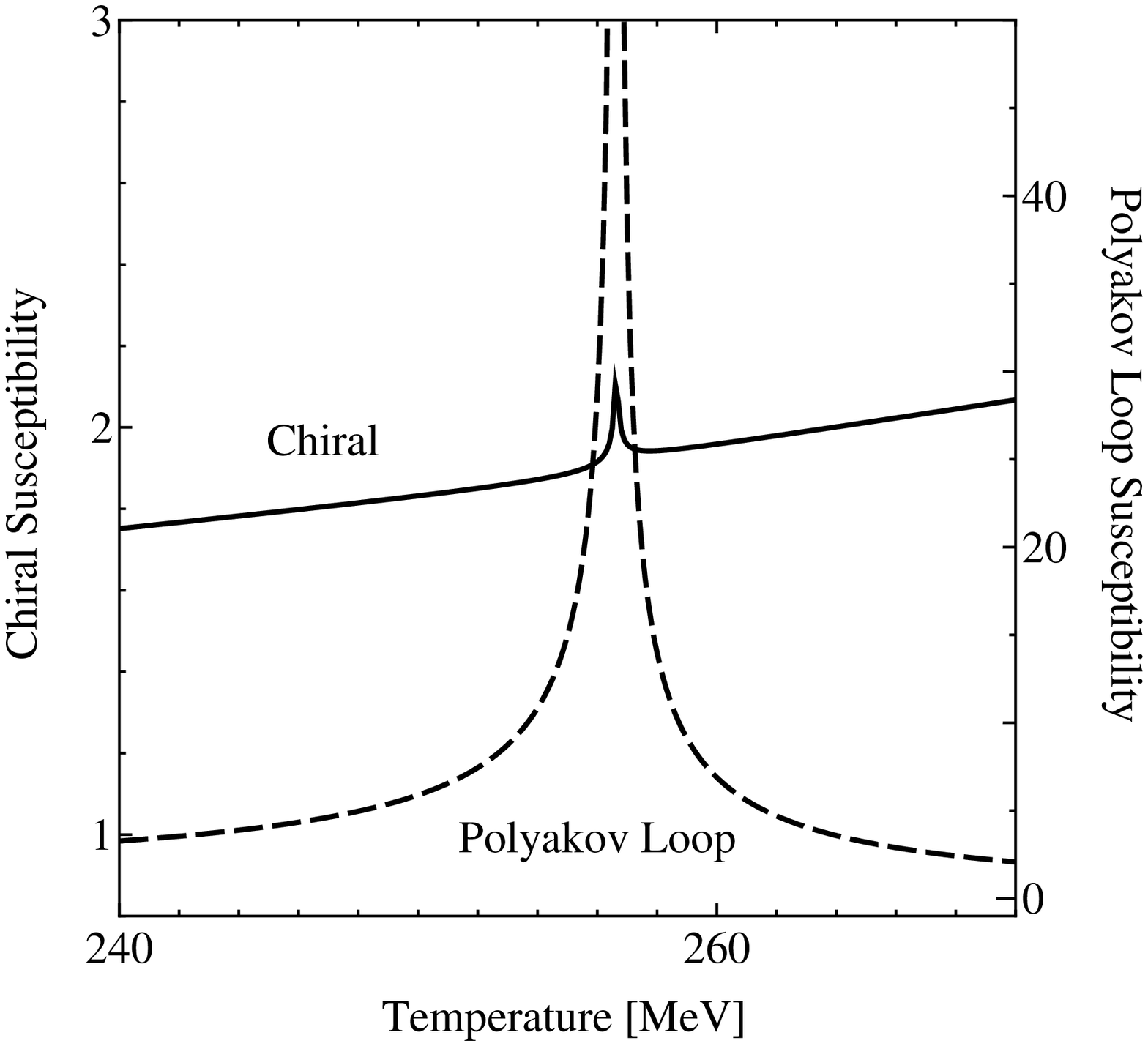}
\caption{The order parameter and susceptibility around the
deconfinement critical end-point with
$m_q^{\textsf{D}}=788\,\text{MeV}$ and $\mu=0\,\text{MeV}$.}
\label{fig:order_dcep}
\end{figure}

At $\mu=0$, as discussed in Introduction, the deconfinement CEP is
important as well as the chiral CEP. [In the present two-flavor case
the chiral CEP trivially lies at $m_q=0$.] We found the deconfinement
CEP at ($m_q^{\textsf{D}}=788\,\text{MeV},\,
T_{\textsf{D}}=257\,\text{MeV}$) in our model. This value is
consistent with the lattice observation in the two-flavor case
\cite{has83} and also in good agreement with the Gocksch-Ogilvie model
\cite{fuk03_2}. The order parameter and the susceptibility around the
deconfinement CEP are shown in Fig.~\ref{fig:order_dcep}. Since the
deconfinement transition is of second-order, the Polyakov loop
susceptibility diverges at $T=T_{\textsf{D}}$. In
Figs.~\ref{fig:order_ccep} and \ref{fig:order_dcep} we can see that
the chiral susceptibility and the Polyakov loop susceptibility both
have a singularity due to the mixing.
\vspace{1mm}


\paragraph*{\bf Discussion}
We shall briefly summarize characteristic features of our model
defined by Eqs.~(\ref{eq:potential}) and (\ref{eq:ansatz}) below.

\begin{enumerate}

\item
The coupling between the Polyakov loop and the chiral condensate is
determined uniquely and is consistent with the conventional form; in
the leading order of the hopping parameter ($\kappa$) expansion on the
lattice, the coupling term takes the form of
$(2\kappa)^{N_\tau}l\sim l\,\mathrm{e}^{-M/T}$ \cite{goc85,gre84} (see
the coupling term of (\ref{eq:potential})).

\item
Because the coupling term is $\sim l\,\mathrm{e}^{-M/T}$, it goes to
zero as $T\to0$. Actually the zero temperature system can be different
from the system at infinitesimally small but finite temperature. At
zero temperature, the canonical description with the quark triality
fixed at zero is likely to be valid \cite{fuk03}.

\item
Contrary to naive expectation, the Polyakov loop behavior hardly
reflects the singularity associated with the chiral phase transition
of second-order. [The gross feature of Fig.~\ref{fig:order} is hardly
changed when $m_q=0$ except that $\chi_q$'s peak becomes divergent.]
This is because the coupling $C_{ql}$ (amplitude between the Polyakov
loop and the chiral condensate) is proportional to the constituent
quark mass, $M$, and vanishes at the chiral phase transition of
second-order. Nevertheless, our idea leads to the almost simultaneous
crossovers in a robust way.

\item
Figure \ref{fig:order_ccep} is an interesting prediction from our
model at finite density. Our idea would not necessarily give the
simultaneous crossovers at high density where $T_{\textsf{E}}$ is too
lower than $T_{\textsf{D}}$ and the Polyakov loop has a long tail.
This double-peak structure with a sharp peak and a broad bump would be
a realistic possibility to be seen in the future lattice simulation at
high density.

\item
As discussed in the section ``Idea'', our idea would not work for
large $m_q$ because of explicit symmetry breaking. In general the
pseudo-critical temperature $T_\chi$ gets larger with increasing
$m_q$. Actually the chiral susceptibility peak calculated in the
standard NJL model yields
$T_\chi=270\;\text{MeV}(\sim T_{\textsf{D}})$ at
$m_q=167\;\text{MeV}$. For $m_q>167\;\text{MeV}$, our model leads to a
double-peak structure with $T_{\text{d}} < T_\chi$, that is not
prohibited by our dynamical mechanism. The chiral susceptibility in
Fig.~\ref{fig:order_dcep} has a second broad bump at much higher
temperature than shown in the figure. Although the present model
embodies our idea and describes the simultaneous crossovers for small
$m_q$, there must be some further constraints, in particular to
impose $T_{\text{d}}\ge T_\chi$. Such a condition would realize a
perfect locking for small $m_q$ and cure the failure for larger $m_q$.

\end{enumerate}


\paragraph*{\bf Summary}
We proposed an idea to realize the simultaneous crossovers of
deconfinement and chiral restoration, which turned out to work well
for small quark mass. We demonstrated the idea by using a chiral
effective model with the Polyakov loop. The model study yields the
chiral CEP at ($\mu_{\textsf{E}}=321\,\text{MeV},\,
T_{\textsf{E}}=106\,\text{MeV}$) and the deconfinement CEP at
($m_q^{\textsf{D}}=788\,\text{MeV},\,T_{\textsf{D}}=257\,\text{MeV}$).
The Polyakov loop susceptibility and the chiral susceptibility both
diverge at the CEP due to the mixing effect.

Since the thermodynamic potential in our model is a function of $M^2$
($M$ being the constituent quark mass), the mixing effect $\propto M$
is small at $T=T_{\textsf{C}}$. Then our idea plays an essential role
to attract one crossover to the other. Also we presented a prediction
from our model at finite baryon density. At sufficiently high density
we can expect that the Polyakov loop has a double-peak structure with
a sharp peak from the mixing and a broad bump from a remnant of
deconfinement.

The present model lacks some mechanism necessary to sustain the
locking for $m_q>167\;\text{MeV}$. In other words, this result
suggests that some dynamical mechanism is further needed in order that
the chiral and deconfinement CEPs are connected by a single crossover
line \cite{hat03}. The scenario of \cite{hat03} requires something
beyond the mixing argument and the present idea to lock two phenomena.
Although this is still an open question, we believe that our model can
contain correct physics at least for small $m_q$ and can be a simple
starting point to examine the underlying relation between
deconfinement and chiral restoration not only at finite temperature
but at finite baryon density also.\vspace{1mm}

\paragraph*{\bf Acknowledgments}
The author, who is supported by Research Fellowships of the Japan
Society for the Promotion of Science for Young Scientists, thanks
R.D.~Pisarski, K.~Rajagopal and O.~Phillipsen for comments. Y.~Hatta
is acknowledged for stimulating discussion. He also thanks
O.~Schr\"{o}der for a careful reading of the manuscript. This work is
supported in part by funds provided by the U.S.\ Department of Energy
(D.O.E.) under cooperative research agreement \#DF-FC02-94ER40818.

\end{document}